\documentclass[final,conference,a4paper]{IEEEtran}

\usepackage[T1]{fontenc}
\usepackage{cite}
\usepackage{url}
\usepackage{amsmath}
\usepackage[cmintegrals]{newtxmath}
\usepackage[caption=false,font=footnotesize]{subfig}
\usepackage{graphicx}
\usepackage{epstopdf}
\usepackage{xcolor}
\usepackage{mathtools}

\newcommand\blfootnote[1]{%
	\begingroup
	\renewcommand\thefootnote{}\footnote{#1}%
	\addtocounter{footnote}{-1}%
	\endgroup
}

\begin{document}
\bstctlcite{IEEEexample:BSTcontrol}

\title{Modeling Profit of Sliced 5G Networks for Advanced Network Resource Management and Slice Implementation}



\author{\IEEEauthorblockN{
  Bin~Han, Shreya~Tayade, and Hans~D.~Schotten}\\
  \IEEEauthorblockA{
  Institute of Wireless Communication and Navigation\\
  Department of Electrical and Computer Engineering\\
  University of Kaiserslautern\\
  67663 Kaiserslautern, Germany\\
  Email: \{binhan, tayade, schotten\}@eit.uni-kl.de}
}

%
%


\maketitle

\begin{abstract}
The core innovation in future 5G cellular networks-network slicing, aims at providing a flexible and efficient framework of network organization and resource management. The revolutionary network architecture based on slices, makes most of the current network cost models obsolete, as they estimate the expenditures in a static manner. In this paper, a novel methodology is proposed, in which a value chain in sliced networks is presented. Based on the proposed value chain, the profits generated by different slices are analyzed, and the task of network resource management is modeled as a multi-objective optimization problem. Setting strong assumptions, this optimization problem is analyzed starting from a simple ideal scenario. By removing the assumptions step-by-step, realistic but complex use cases are approached. Through this progressive analysis, technical challenges in slice implementation and network optimization are investigated under different scenarios. For each challenge, some potentially available solutions are suggested, and likely applications are also discussed.

\end{abstract}

\begin{IEEEkeywords}
network slicing, cost model, 5G network, network optimization, resource management
\end{IEEEkeywords}

\IEEEpeerreviewmaketitle

\section{Introduction}
\blfootnote{This is a preprint, the full paper has been published in the 2017 IEEE Symposium on Computers and Communications (ISCC), \copyright 2017 IEEE. Personal use of this material is permitted. However, permission to use this material for any other purposes must be obtained from the IEEE by sending a request to pubs-permissions@ieee.org.}

Higher system scalability and flexibility, calls for better energy, cost and resource utilization efficiencies in future 5G mobile communication networks\cite{osseiran2013foundation, osseiran2014scenarios}. As an architectural answer to this, network slicing is considered as the most important and innovative concept in 5G, and has been intensively discussed in recent years\cite{ericsson2015five,ngmn2015deliverable,nikaein2015network}. With the idea of organizing physical networks in the form of logically separate and independent slices, network slicing enables a flexible and polymorphic network customization according to the specific use case. This concept is changing the network configuration process. Instead of generally optimizing the entire network for all customers and services, the operators are now able to separately optimize each slice with an individual configuration. 
To implement network slices efficiently
, a cost function must be constructed for each slice. Classical mobile network cost models that are based on long-term capital expenditures (CAPEX) and operating expenditures (OPEX), as in 
\cite{knoll2014combined} and \cite{nikolikj2014cost}, cannot support modeling costs of logical network slices. Till now, no appropriate cost model for sliced networks has been proposed. In this paper, we propose a novel methodology of analyzing and optimizing the business profit generated by a network slice. Through the discussions under different cases, we try to reveal the key challenges in slice implementation, and suggest several candidate solutions to them. The paper is structured as follows. In Sec. \ref{sec:intro_ns} we briefly introduce the concept of network slicing. Then in Sec. \ref{sec:slice_profit} we propose a novel profit model, which maps the slice properties, such as key performance indicator (KPI) requirements and size, to the profit generated by the slice. Based on this model, in Sec. \ref{sec:slice_optimization} we analyze the sliced network optimization problem, so as to maximize the profits of network operator. We set several strong assumptions to initiate the analysis under simple and ideal conditions, and progressively approach to the complex realistic case by removing the assumptions step-by-step. By the end we close this paper with Sec. \ref{sec:conclusion}, where some conclusion and outlooks to future studies are given.

\section{Network Slicing for Flexibility and Efficiency}\label{sec:intro_ns}
\IEEEpubidadjcol
The modern mobile communication technologies have been providing various types of services, such as video streaming, mobile cloud storage, mobile online gaming, etc. Nevertheless, the quality of service (QoS) is usually limited in existing cellular networks. Consider a common scenario: a user is playing an online game with his mobile phone, while listening to online music with the same device. Meanwhile, the mobile phone is downloading a large file from a cloud storage server. Here, three data services are sharing the same radio access network (RAN). Due to the limitations in the network resources, the user may suffer from an intermittent music, a significant delay in the game interaction and a low speed in downloading a file. 

However, the lack of resources is not the real problem. As a matter of fact, purpose-built applications usually have specific performance demands that highly depend on the use case. 
For example, the online music streaming service can tolerate a relatively high buffering time at the the beginning, but it requires a certain level of data rate with high availability and retainability to promise the continuity of music. 
In contrast, online games commonly require low latency, high retainability and high reliability to guarantee a smooth game experience, despite their low data size.
Cloud storage synchronization, differing from them both, usually asks for a high channel capacity, while having lenient requirements on latency. 
Most modern cellular networks possess enough resources to fulfill the demands of an arbitrarily selected service, especially when optimizing the network resources correspondingly.
Nevertheless, the current network architecture is not capable to support all new kinds of services and scenario. Therefore, the aforementioned scenario, can result in a regrettable QoS, regardless of the richness in network resources. Furthermore, to enhance the performance in one particular use case, a network usually has to be solely scaled up, which often implies an exorbitant investment.

\begin{figure}
	\centering
	\includegraphics[width=.45\textwidth]{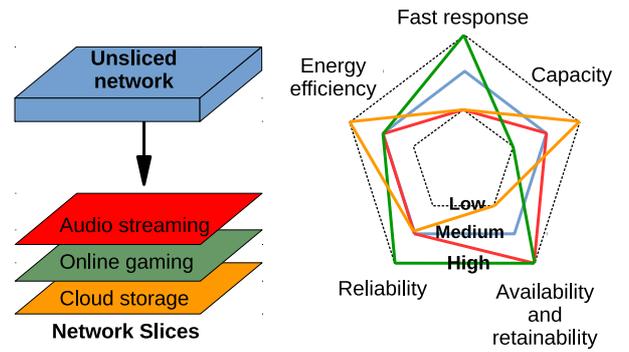}
	\caption{Slicing an omni-functional network into three specialized network slices}
	\label{fig:network_slicing}
\end{figure}

As one of the key enablers of future 5G networks, network slicing can help solve this problem with a much more flexible and efficient resource allocation. A network slice is an abstracted connectivity service, i.e. a logical network, which is defined by a number of customizable software-defined functions\cite{ericsson2015five}. Resources of a physical network can be logically allocated to different slices. According to the performance demands, various use cases can be categorized into several services. For each service, a network slice can be abstracted and a software-defined implementation can be optimized. With respect to the requirement, each slice can be separately scaled, in order to adjust the corresponding service performance, independently of other slices. Hence, the network configuration can be executed in the form of slice size configuration. In this way, the network resources can be flexibly organized and efficiently managed, to fulfill different demands of various services. An example is presented in Fig. \ref{fig:network_slicing}, by which three slices are defined and optimized for audio-streaming, online gaming and could storage, respectively.

\section{The Slice Business Model:\\from Slice Properties to Profit}\label{sec:slice_profit}
Constrained by the overall resource limits, the network has to compromise among the sizes of different slices, which leads to an optimization problem. According to the network operators, it is usually the business profit to maximize. To achieve this, a value chain in the network slicing process must be built.

Classical network cost models are typically built based on CAPEX and OPEX, which are estimated according to the number of base stations (BSs), the transmission power and the traffic volume \cite{knoll2014combined,nikolikj2014cost}. For sliced networks, this methodology is not appropriate any more. As each resource can be shared by several network slices, and the slicing scheme varies from one resource to another. Hence, OPEX cannot be generally estimated for the entire physical network. A novel slice-oriented cost model is therefore needed.

\begin{figure}
	\centering
	\includegraphics[width=.45\textwidth]{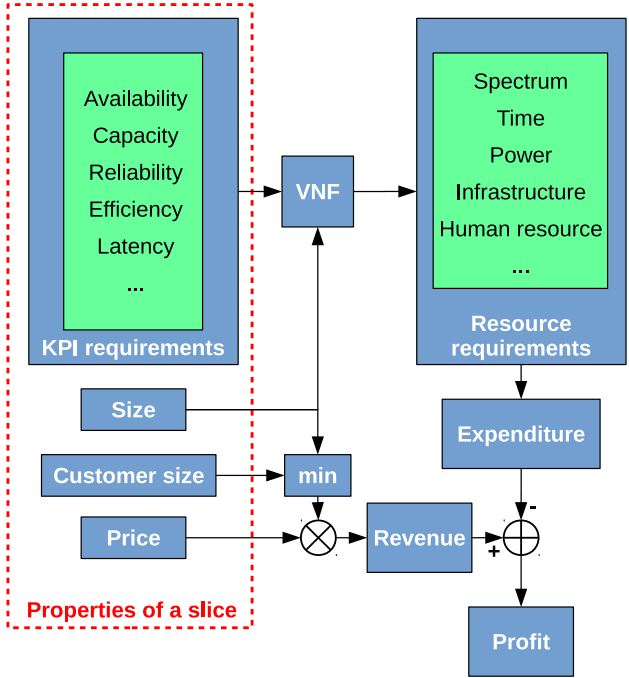}
	\caption{A mapping chain that converts the properties of a slice into the slice profit step-by-step}
	\label{fig:mapping_chain}
\end{figure}

As we already discussed, every slice is specified for a particular pre-defined service, which covers a group of use cases with similar demands for the QoS. Therefore, a slice can be identically defined by a set of KPI requirements, as referred to in many 5G research projects \cite{metis2015deliverable,5gppp2016view}.
The complete mapping chain from slice requirement to operator's profit is illustrated in Fig. \ref{fig:mapping_chain}.
Given a set of KPI requirements $\mathbf{k}=[k_1,k_2,\dots,k_L]$, a virtual network function (VNF) can be designed in the specification phase to achieve them. According to the VNF implementation $\mathcal{V}$ and the slice size $s$ (the maximal number of user applications that can be served by the slice), the required volume of network resources can be estimated. Various kinds of resources can be enumerated, e.g. spectrum/bandwidth, time, power, infrastructures, human resource etc. Let us record the required amount of them in a vector $\mathbf{r}=[r_1,r_2,\dots,r_N]$, where $N$ is the number of resource types. According to the cost of each resource, the resource requirements can be further converted into the expenditure $\mathrm{EXP}$, in a similar way as in classical network cost models. So that we have
\begin{align}
\mathrm{EXP} &= \mathrm{EXP}(\mathbf{r}),\\
\mathbf{r} &= \mathbf{r}(\mathbf{k},s,\mathcal{V}).
\end{align}
On the demand side, meanwhile, a certain price $p$ must be paid by the network customers for each service. Therefore, given the service price, the slice size $s$ and the customer size $c$ (the number of user applications requesting service from the slice), the revenue of a slice can be straightforwardly modeled as
\begin{equation}
\mathrm{REV} = \mathrm{REV}(p,s,c).
\end{equation}
By subtracting the cost from the revenue, the profit generated by the slice can be computed as
\begin{equation}
w=\mathrm{REV}(p,s,c)-\mathrm{EXP}(\mathbf{r}) = w(\mathbf{r},p,s,c).
\end{equation} 

It is worth to note, that the KPI-to-resource mapping $\mathbf{r}(\mathbf{k},s,\mathcal{V})$ is not only 
very complex, but also highly dependent on the selection of VNF implementation $\mathcal{V}$. 
Nevertheless, as the network operator is responsible for the VNF implementation, it always holds a full knowledge about it. Therefore, in the operator's point of view, it is reasonable to assume the function $\mathbf{r}(\mathbf{k},s,\mathcal{V})$ as a-priori known.

\section{Network Optimization: Maximizing the Profits}\label{sec:slice_optimization}
In a network with multiple slices $i=1,2,\dots,M$, the operator aims to maximize the profits $w_i$ of all $M$ slices simultaneously by allocating $N$ different resources to them. This is a multi-objective optimization problem (MOOP):
\begin{align}\label{equ:MOOP}
\begin{split}
\arg\max\limits_\mathbf{R}&\mathbf{w}(\mathbf{R}),\\ \mathrm{where}\quad\mathbf{w}(\mathbf{R})&=[w_1(\mathbf{R}),w_2(\mathbf{R}),\dots,w_M(\mathbf{R})]^\mathrm{T},\\
\mathbf{R}&=[\mathbf{r}_1,\mathbf{r}_2,\dots,\mathbf{r}_M]^\mathrm{T},\\
\mathbf{r}_i&=[r_{i,1},r_{i,2},\dots r_{i,N}].
\end{split}
\end{align}
Here, $r_{i,j}$ indicates the volume of the $j^\mathrm{th}$ resource allocated to the $i^\mathrm{th}$ slice.
The optimization is constrained by two kinds of boundary conditions:
\begin{itemize}
	\item The utilized resource bundle is physically limited by the total volume of resource pool $\mathbf{r}_\Sigma$.
	\item The lower boundaries of slice sizes are manually established according to the network service policies. So that each slice $i$ must at least be provided with a pre-defined minimum level of resources $\mathbf{r}^\mathrm{min}_i$, even if some slices appear commercially unattractive to the operator.
\end{itemize}

To study this problem progressively, in this section we first simplify it with some strong assumptions on the use scenario and the VNF implementation, then we relax the assumptions step-by-step to approach the complex realistic use cases.

\subsection{Start with the Ideal Model: Assumptions}
For a strong simplification, we set the following assumptions:
\begin{enumerate}
	\item \textit{Resource shortage}: the user demands are heavy while the network resources are limited, so that the network cannot satisfy all the service demands simultaneously.
	\item \textit{Single operator}: no multi-operative service is supported, i.e. the network is operated by a single operator, which holds the complete network information.\label{ass:single_operator}
	\item \textit{Static slices}: the KPI requirements, the customer size and the service price of every slice remain constant.\label{ass:static_slices}
	\item \textit{Open-loop system}: the configuration of network slices does not have any unexpected short-term impact on the network environment.
	\label{ass:open_loop_system}
	\item \textit{Orthogonal slices}: all the slices are orthogonal to each other, i.e. no resource can be shared by two or more different slices, and the VNF implementation of a slice only depends on its own KPI requirements and size.\label{ass:independent_slices}
\end{enumerate}
These assumptions together set up a classical Multi-Objective Optimization Problem (MOOP), where the single network operator tries to maximize the profits $w_i$ made by all slices $i=1,2,\dots,N$ simultaneously. 

A variety of methods have been developed to solve MOOPs \cite{bjornson2014multiobjective,marler2004survey,sunar2001comparative}.
The most common one of them is the weighted sum method, which transforms the target MOOP (\ref{equ:MOOP}) into a conventional single-objective optimization problem (SOOP):
\begin{equation}
\arg\max\limits_\mathbf{R}\sum\limits_{i=1}^{M}g_iw_i(\mathbf{r}_i,s_i,c_i,p_i),
\label{equ:SOOP_weighted_sum}
\end{equation}
where $s_i$, $c_i$ and $p_i$ are the size, the customer size and the service price of slice $i$. $g_i$ is a weight factor, which represents the importance preference of slice $i$. If the operator has no special preference for any slice, we can take $g_i=1$ for all $i$, to maximize the overall profit:
\begin{equation}
\arg\max\limits_\mathbf{R}\sum\limits_{i=1}^{M}w_i(\mathbf{r}_i,s_i,c_i,p_i),
\label{equ:SOOP_objective_sum}
\end{equation}	
which is known as the objective sum method.

As mentioned before, the resources $\mathbf{r}_i$ needed by slice $i$ are determined by the corresponding KPI requirements $\mathbf{k}_i$, $s_i$ and the VNF implementation $\mathcal{V}_i$, i.e.
\begin{align}
\begin{split}
\mathbf{r}_i &= \mathbf{r}_i(\mathbf{k}_i,s_i,\mathcal{V}_i),\\
\mathrm{where}\quad\mathbf{k}_i &=[k_{i,1},k_{i,2},\dots,k_{i,L}].
\end{split}
\end{align}
Here, $k_{i,j}$ indicates the $j^\mathrm{th}$ KPI requirement of the $i^\mathrm{th}$ slice and $L$ the dimension of the KPI requirements.
Under the assumption 
of orthogonal slices, $\mathcal{V}_i$ is a function of $\mathbf{k}_i$ and $s_i$. Hence, $w_i(\mathbf{r}_i,s_i,p_i)$ can be represented as $w_i(\mathbf{k}_i, s_i, p_i)$. Furthermore, the static slices assumption 
suggests that $\mathbf{k}_i$, $c_i$ and $p_i$ are constants, so that both $\mathbf{r}_i$ and $w_i$ can be considered as functions of $s_i$. Hence, the SOOP (\ref{equ:SOOP_objective_sum}) turns into
\begin{align}
\begin{split}
&\arg\max\limits_\mathbf{s}\sum\limits_{i=1}^{M}w_i(s_i),\\
\mathrm{where}\quad&\mathbf{s}=[s_1,s_2,\dots,s_M]^\mathrm{T}.
\end{split}
\label{equ:SOOP_simplest}
\end{align}
And the boundary constrains can be represented as:
\begin{align}
\sum\limits_{i=1}^{M}r_{i,j}&\le r_{\Sigma,j},\quad j=1,2,\dots,N;\label{equ:linear_resource_pool_boundaries}\\
r_{i,j}&\ge r_{i,j}^\textrm{min},\quad i=1,2,\dots,M,\quad j=1,2,\dots,N, \label{equ:minimal_slice_sizes}
\end{align}
where $r_{\Sigma,j}$ represents the total volume of resource $j$ and $ r_{i,j}^\textrm{min}$ the minimal volume of resource $j$ reserved for slice $i$.

\subsection{Sharing for Good: Slice Multiplexing}
In real implementation, different slices are usually sharing some resources to achieve multiplexing, if not always. For example, a massive Machine-Type-Communication (mMTC) slice and an ultra-reliable MTC (uMTC) slice can share the same base stations, some network functions such as radio scheduler, or even the same band in spectrum (under time division). In this case, one slice may benefit from a size decrease 
of another slice. Furthermore, to achieve a higher multiplexing gain, the multiplexing scheme should be carefully planned before the network function virtualization (NFV), which impacts the VNF implementation of every slice. To reveal these effects, we remove the assumption of orthogonal slices
from our list.

\begin{figure}
	\centering
	\includegraphics[width=.5\textwidth]{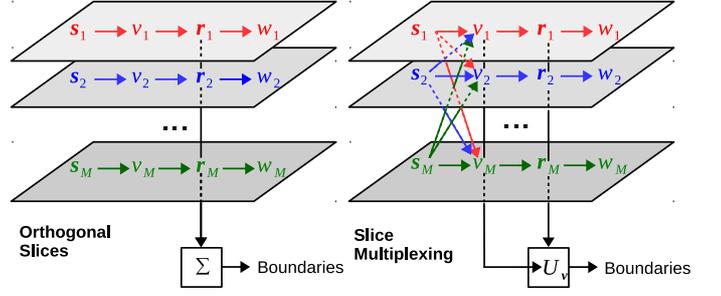}
	\caption{Slice multiplexing dramatically complicates the mapping chain for profit optimization. Under multiplex, the size of each slice impacts the VNF implementations of all the other slices that are multiplexed with it, and the total resource cost is a VNF-dependent function of different slice resource costs.}
	\label{fig:slice_multiplexing_complexity}
\end{figure}

This complicates the problem dramatically in two aspects. First, the SOOP simplification from (\ref{equ:SOOP_objective_sum}) to (\ref{equ:SOOP_simplest}) is no more valid, with the assumption $\mathcal{V}_i=\mathcal{V}_i(\mathbf{k}_i,s_i)$ removed. Instead, due to the multiplexing scheme selection stage, the implementation $\mathcal{V}_i$ of every slice $i$ must be jointly selected with the others $\mathcal{V}_{j, j\neq i}$, according to the global network requirements, i.e.
\begin{align}
\begin{split}
\mathcal{V}_i &= \mathcal{V}_i(\mathbf{K},\mathbf{s}),\\
\mathrm{where}\quad\mathbf{K}&=[\mathbf{k}_1,\mathbf{k}_2,\dots,\mathbf{k}_M]^\mathrm{T}.
\end{split}
\end{align}
Hence, the SOOP (\ref{equ:SOOP_objective_sum}) turns into
\begin{equation}
\arg\max\limits_{\mathbf{s}}\sum\limits_{i=1}^{M}w_i(\mathbf{s}),
\label{equ:SOOP_multiplex}
\end{equation}
In (\ref{equ:SOOP_simplest}), the profit function $\mathbf{w}(\mathbf{s})$ can be decomposed into $M$ independent functions $w_i(s_i)$, each of only one variable. Compared to the simple case, the profit function under slice multiplexing can only be jointly optimized with respect to $M$ variables. Second, besides the slice sizes, the volume of utilized resources strongly depends on the multiplexing mode, i.e. the VNF implementations. Hence, instead of the simple linear function (\ref{equ:linear_resource_pool_boundaries}), the resource pool boundaries must be described by
\begin{equation}
\mathcal{U}_\mathbf{V}(r_{1,j},r_{2,j},\dots,r_{M,j})\le r_{\Sigma,j},\quad j=1,2,\dots,N,\label{equ:flexible_resource_pool_boundaries}
\end{equation}
where $\mathcal{U}_\mathbf{V}$ is a function with form depending on the VNF implementations $\mathbf{V}=[\mathcal{V}_1,\mathcal{V}_2,\dots,\mathcal{V}_M]^\mathrm{T}$. Fig. \ref{fig:slice_multiplexing_complexity} briefly illustrates these effects.

Certainly, the operator still holds full knowledge about the VNF implementation, so that the SOOP (\ref{equ:SOOP_multiplex}) can still be solved in a centralized approach, and a Pareto optimum is theoretically available. However, the high ranks of the cost function (\ref{equ:SOOP_multiplex}) and the boundary constraints (\ref{equ:flexible_resource_pool_boundaries}) can significantly raise the computational cost, probably to a level beyond an operator could afford.

To reduce the computational cost, one simple solution is to limit the number of available slice-multiplexing modes, i.e. the candidate VNF implementations $\mathbf{V}$. The candidate set should be small enough to traverse, and carefully selected, to guarantee a reasonable efficiency. As a trade-off, the slice multiplexing gain will be reduced. Besides, it is doubtful if any universal approach is available to define such a candidate VNF set.

Another possible approach would be to decompose the multi-dimensional optimization into two separate levels and solve it iteratively. First, the VNF scheme is initialized to a blueprint, under which the slice sizes are optimized. Next, the slice sizes are fixed at the optimization result, while the VNFs are reimplemented to further improve the network profit. These two steps repeat recursively until the network configuration converges, or after a given threshold of iteration loops. This approach follows the idea of Block Coordinate Descent (BCD) method \cite{tseng2001convergence}, which is widely used in MOOPs. However, differing from standard applications of BCD methods, the VNF implementations $\mathcal{V}_i$ are not variables of real value, but parameters that determine the mathematical representation of the cost function. This nature of $\mathbf{V}$ can significantly impact the converging ability of the method, which must be investigated before any application.

Considering the absence of gradient information with respect to $\mathbf{V}$, and our aim at global optimum instead of local ones, genetic algorithms (GAs) exhibit an ubiquitous advantage in exploring the Pareto front of the SOOP (\ref{equ:SOOP_multiplex}) \cite{marler2004survey}. The basic idea of genetic algorithms is to generate new solutions from an initial solution population, in a recursive and evolutionary manner. According to the fitness value, which can be computed for every solution to evaluate its degree of optimum, different solutions have their individual opportunities to produce posterities
in further iterations. After enough number of generations, the remaining solution population is expected to converge at the Pareto front. This class of algorithms are able to search the solution space in multiple and random directions in every single iteration, and hence show high efficiency in solving MOOPs with large degree of freedom, such as the slice optimization. Detailed tutorials of applying GA in MOOPs can be found in \cite{deb2001multi} and \cite{coello2007evolutionary}.

\subsection{Environment Responses: Iterative Optimization}
Permitting different slices to multiplex benefits with a multiplexing gain, but it also leads to a much more complex network environment. As long as slices are allowed to share spectral and time resources, they are no more physically isolated from each other. Instead, each slice can become part of the environment for other slices. For example, if the mMTC and uMTC slices share the same spectrum and time by using different BS antennas or by applying code division, when we raise the transmission power of mMTC signals, the interference strength is increased for uMTC. Therefore, the slice configuration now impacts the network environment, so the network is a closed-loop system where $\mathbf{K}=\mathbf{K}(\mathbf{R})$. Hence the assumption of open-loop system 
should be removed.

Practically, there are plenty potential sources of such inter-slice impacts, that cannot be completely known beforehand but can only be measured posteriorly. In this case, the optimum can only be solved in iterative approaches. An initial optimization should first be executed under given KPI requirements $\mathbf{K}=\mathbf{K}_0$. Then network measures the environment after applying the optimization, and respectively update the KPI requirements to $\mathbf{K}=\mathbf{K}_1$. Afterwards, the network optimization is repeated under the updated $\mathbf{K}$. This process continues iterating, until it converges.

\subsection{Optimization Validity versus Configuration Cost: Long-Term Optimization}
Besides the short-term variations caused by the network configuration, the environment also exhibits middle-term and long-term fluctuations in nature. For example, for a given slice $i$, the customer size $c_i$ often keeps varying through a day, because the user activity highly depends on time. The channel conditions also usually change due to the unstable electro-magnetic environment, which leads to fluctuating KPI requirements $\mathbf{k}_i$. Even the service price $p_i$ can be occasionally adjusted by the operator for business reasons as well. These facts suggest to cancel our assumption of static slices
, and to keep dynamically updating the slices in real time.

However, it must be considered that the updating process of slice configuration generates extra cost itself. To accomplish such an update, lot of measurements and signaling are essential, which cost network resources, especially power. So far, we are now facing a new compromise: with a shorter period of slice update, the profit loss caused by the outdated slice configuration can be reduced, while the reconfiguration cost increases; and vice versa. To achieve an ideal balance, a long-term optimization respecting to the update period should be implemented, based on the reconfiguration cost and the historical short-term optimization results.

\subsection{Multi-operator and Sub-operator: Game Theoretic Approaches}
So far, we have been working with the slice scaling optimization, in order to maximize the resource utilization efficiency and the overall profit within a network owned by a single operator. Yet, such an optimization cannot guarantee a full resource utilization. Even after a complete optimization there are usually network resources that still remain idle and cannot support additional slice implementation. On the other hand, there are also some use cases with extremely specialized KPI requirements, which can be hardly fulfilled by any single operator, even if a plenty of resources are available. A typical example is the Public Protection and Disaster Relief (PPDR), which demands very low data rate but ultra-high availability. Although the service availability can be improved in some order by increasing the transmission power, the only effective approach to achieve a significant enhancement in availability, as the PPDR demands, is to deploy more access points. It leads to a huge CAPEX, if a network operator has to invest all the infrastructures for new APs by itself. Meanwhile, the revenue generated by the PPDR service may be quite limited, so far as it only utilizes a very narrow band in the spectrum.

Therefore, multi-operative service can be a win-win solution. In the multi-operative framework, several network operators are able to trade their idle resources with each other for certain prices. For example, an operator A can rent an under-utilized small band in the spectrum at all its base stations to another operator B, in order to support an ultra-available PPDR service of operator B. In this way, operator B can extend its resource pool with slight increase in OPEX but without any increase in CAPEX. Also, operator A can now generate an additional profit by sharing idle resources. The multi-operator framework, hence significantly increases the overall network resource utilization factor.

Nevertheless, to realize multi-operative services, the prices of resources in trade must be reasonably and flexibly decided. This is also a MOOP, where the profit of every operator gained from the trading is to maximize. However, differing from the cases above, no SOOP can be generated here to solve it, because the operators are not sharing their internal business information such as slice size or resource cost. Hence, in absence of global system knowledge, no centralized decision making mechanism is available.

This condition, where different operators make their own decisions in purpose of profit maximization, self-evidently encourages to apply game theory (GT) methods. A simple mechanism, for example, is to let the operators offer their prices for their lacked/idle resources in a transparent market. According to the available market information, every operator makes rational trading decisions for its own profit. Generally, in such kind of scenarios, the game is non-cooperative and with imperfect information.

Another possible multi-operative scheme is that, one network operator can run its business through several sub-operators, each one is responsible for a different slice. The main operator owns all network resources and dynamically allocate them to its sub-operators. In this way, a game is settled where the sub-operators compete for network resources as individual players. Differing from the case discussed above, as all the sub-operators belong to the same operator, the game can be designed as cooperative and with perfect information.

GT methods helps to make decisions in a decentralized way, which enables multi-operative services, and reduces the computational complexity of network optimization in the sub-operator case. However, when applying GT methods, it is critical to distinguish the Pareto optimum, which is efficient and expected, from the Nash equilibrium, which is only strategical feasible and usually obtained. A Nash equilibrium is not necessarily a Pareto optimum.

\section{Conclusion and Outlooks}\label{sec:conclusion}
In this paper, we have analyzed the impact of network slicing on resource management and profit generation. We have proposed a novel methodology of modeling profit generated by 5G network slices. The expenditure and revenue of a network can be estimated according to its slice properties, such as KPI requirements and service prices. Based on this, the slice implementations can be optimized to maximize the profit . Through case studies under different assumptions, we have set a MOOP model for the sliced network optimization, and revealed the technical challenges in this task. For each challenge, we have suggested some candidate approaches for eventual solution.

In future, the proposed framework can be extended in various aspects, including but not limited to the following ones. The contents of KPI and resource requirements should be well defined to achieve a possibly simple and clear mapping. Realistic models of resource cost and service revenue with detailed parameters shall be developed and integrated into the proposed framework, in order to support quantitative simulations for evaluation of the suggested solutions. As the final target, the most appropriate solutions can be implemented, to overcome the technical challenges and to realize the envisaged network optimization.

\section*{Acknowledgment}
This work has been performed in the framework of the H2020-ICT-2014-2 project 5G NORMA. The authors would like to acknowledge the contributions of their colleagues. This information reflects the consortium's view, but the consortium is not liable for any use that may be made of any of the information contained therein.


\bibliographystyle{IEEEtran}
\bibliography{references}

\begin{IEEEbiographynophoto}{Bin Han}
received his B.E. degree in 2009 from the Shanghai Jiao Tong University, China, and his M.Sc. degree in 2012 from the Technische Universit\"at Darmstadt, Germany. He was granted the Ph.D. (Dr.-Ing.) degree in Electrical Engineering and Information Technology from the Kalsruhe Institute of Technology, Germany, in 2016. He is currently a postdoctoral researcher at the Institute of Wireless Communication and Navigation (WiCoN), in the Department of Electrical and Computer Engineering, University of Kaiserslautern, Germany. His research interests are in the broad area of communication systems and signal processing. In the context of wireless communication, his interests include massive Machine-Type-Communications and 5G cellular networks.
\end{IEEEbiographynophoto}

\begin{IEEEbiographynophoto}{Shreya Tayade}
received her B.E from Government College of Engineering Aurangabad, India in 2010 and M.Sc degree in Communication Engineering from RWTH Aachen University, Germany in 2015. Since 2015, she is a researcher in University of Kaiserslautern, Germany, and is working on 5G based EU and industrial projects.
\end{IEEEbiographynophoto}

\begin{IEEEbiographynophoto}{Hans D. Schotten}
received the Diploma and Ph.D. degrees in Electrical Engineering from the Aachen University of Technology RWTH, Germany in 1990 and 1997, respectively. He held positions as senior researcher, project manager, and head of research groups at the Aachen University of Technology, Ericsson Corporate Research, and Qualcomm Corporate R\&D. At Qualcomm he was Director for Technical Standards and research coordinator for Qualcomm's participation in national and European research programs. He was company representative at WWRF, EICTA, bmco forum and European technology platforms. Since August 2007, he has been full professor and head of the Institute of Wireless Communication and Navigation at the University of Kaiserslautern. Since 2012, he has additionally been Scientific Director at the German Research Center for Artificial Intelligence heading the "Intelligent Networks" department.
\end{IEEEbiographynophoto}

\end{document}